\documentclass[12pt]{emulateapj}  
\usepackage{graphics}
\usepackage{longtable}
\usepackage{multirow}
\usepackage{rotating}
\usepackage{natbib}
\usepackage{amsmath}
\usepackage{mathrsfs}
\usepackage{eucal}
\usepackage{txfonts}
\newcommand{\kms}{km\,s$^{-1}$}       
\newcommand{\cmtwo}{cm$^{-2}$}

\newcommand{\um}{$\mu$m}                                 
\newcommand{\msun}{M$_{\odot}$}

\newcommand{\lsim}{\;\lower.6ex\hbox{$\sim$}\kern-7.75pt\raise.65ex\hbox{$<$}\;}
\newcommand{\gsim}{\;\lower.6ex\hbox{$\sim$}\kern-7.75pt\raise.65ex\hbox{$>$}\;}
\newcommand{\gl}{\;\lower.6ex\hbox{$<$}\kern-7.75pt\raise.65ex\hbox {$>$}\;}
\newcommand{\amin}{$^{\prime}$}                   
\newcommand{\asec}{$^{\prime \prime}$}
\newcommand{\adeg}{$^{\circ}$}

\newcommand{\sgra}{Sgr A$^{\star}$}
\newcommand{\sgrb}{Sgr B2}
\newcommand{\sgrc}{Sgr C}
\newcommand{\lmbp}{[$l^-,b^+$]}
\newcommand{\lmbm}{[$l^-,b^-$]}
\newcommand{\lpbp}{[$l^+,b^+$]}
\newcommand{\lpbm}{[$l^+,b^-$]}

\shorttitle{The \textit{Herschel} view of the Galactic Center}
\shortauthors{Molinari et al.}

\begin{document}
\title{A 100-parsec elliptical and twisted ring of cold and dense molecular clouds revealed by \textit{Herschel}\footnotemark[*] around the Galactic Center.} \footnotetext[*]{{\it Herschel} is an ESA space observatory with science instruments provided by European-led Principal Investigator consortia and with important participation from NASA.}

\author{S. Molinari\altaffilmark{1}, J. Bally\altaffilmark{2}, A. Noriega-Crespo\altaffilmark{3}, M. Compi\`{e}gne\altaffilmark{3}, J. P. Bernard\altaffilmark{4}, D. Paradis\altaffilmark{4}, P. Martin\altaffilmark{5}, L. Testi \altaffilmark{6, 7}, M. Barlow\altaffilmark{8}, T. Moore\altaffilmark{9}, R. Plume\altaffilmark{10}, B. Swinyard\altaffilmark{8, 11}, A. Zavagno\altaffilmark{12}, L. Calzoletti\altaffilmark{13}, A. M. Di Giorgio\altaffilmark{1}, D. Elia\altaffilmark{1}, F. Faustini\altaffilmark{13}, P. Natoli\altaffilmark{14}, M. Pestalozzi\altaffilmark{1}, S. Pezzuto\altaffilmark{1}, F. Piacentini\altaffilmark{15}, G. Polenta\altaffilmark{13}, D. Polychroni\altaffilmark{1}, E. Schisano\altaffilmark{1}, A. Traficante\altaffilmark{14}, M. Veneziani\altaffilmark{15}, C. Battersby\altaffilmark{2}, M. Burton\altaffilmark{16}, S. Carey\altaffilmark{3}, Y. Fukui\altaffilmark{17}, J. Z. Li\altaffilmark{18}, S. D. Lord\altaffilmark{19}, L. Morgan\altaffilmark{9}, F. Motte\altaffilmark{20}, F. Schuller\altaffilmark{21}, G. S. Stringfellow\altaffilmark{2}, J. C. Tan\altaffilmark{22}, M. A. Thompson\altaffilmark{23}, D. Ward-Thompson\altaffilmark{24}, G. White\altaffilmark{11, 25}, G. Umana\altaffilmark{26}}

\altaffiltext{1}{INAF-IFSI, Via Fosso del Cavaliere 100, I-00133 Roma, Italy}
\altaffiltext{2}{CASA, University of Colorado, Boulder CO, USA 80309}
\altaffiltext{3}{Spitzer Science Center, Caltech, 91125 Pasadena, USA}
\altaffiltext{4}{CNRS, IRAP, F-31028 Toulouse cedex 4,  France}
\altaffiltext{5}{Department of Astron. \& Astroph., University of Toronto, Canada}
\altaffiltext{6}{ESO Headquarters, Garching, Germany}
\altaffiltext{7}{INAF-Osservatorio Astrofisico di Arcetri, 50125 Firenze, Italy}
\altaffiltext{8}{Department of Physics and Astronomy, UCL, London, UK}
\altaffiltext{9}{Astrophysics Research Institute, Liverpool John Moores University, UK}
\altaffiltext{10}{Department of Physics \& Astronomy, University of Calgary, Canada}
\altaffiltext{11}{STFC, Rutherford Appleton Labs, Didcot, UK}
\altaffiltext{12}{LAM, Universit\'{e} de Provence, Marseille, France}
\altaffiltext{13}{ASI Science Data Center, I-00044 Frascati, Italy}
\altaffiltext{14}{Dipartimento di Fisica, Universit\`a di Roma "Tor Vergata", Roma, Italy}
\altaffiltext{15}{Dipartimento di Fisica, Universit\`a di Roma "La Sapienza", Roma, Italy}
\altaffiltext{16}{School of Physics, University of New South Wales, Australia}
\altaffiltext{17}{Department of Astrophysics, Nagoya University, Nagoya, Japan}
\altaffiltext{18}{NAO, Chinese Academy of Sciences, Beijing, China}
\altaffiltext{19}{NASA Herschel Science Center, Caltech, 91125 Pasadena, USA}
\altaffiltext{20}{Laboratoire AIM, CEA/DSM - INSU/CNRS - Universit\'e Paris Diderot, IRFU/SAp CEA-Saclay, 91191 Gif-sur-Yvette, France}
\altaffiltext{21}{MPIfR-MPG, Bonn, Germany}
\altaffiltext{22}{Departments of Astronomy \& Physics, University of Florida, Gainesville, FL 32611, USA}
\altaffiltext{23}{Centre for Astrophysics Research, Science and Technology Research Institute, University of Hertfordshire, Hatfield, UK}
\altaffiltext{24}{School of Physics and Astronomy, Cardiff University, Cardiff, UK}
\altaffiltext{25}{Department of Physics and Astronomy, The Open University, Milton Keynes, UK}
\altaffiltext{26}{INAF-Osservatorio Astrofisico di Catania, Catania, Italy}

\begin{abstract}
Thermal images of  cold dust in the Central Molecular Zone of the 
Milky Way,  obtained with the far-infrared cameras on-board the 
Herschel satellite,  reveal a $\sim 3\times 10^7$ \msun\  ring 
of dense and cold clouds orbiting the Galactic Center. Using a simple 
toy-model, an elliptical shape having semi-major axes of 100 and 60 
parsecs is deduced. The major axis of this 100-pc ring is inclined 
by about 40\adeg\ with respect to the plane-of-the-sky and is oriented 
perpendicular  to the major axes of the Galactic Bar.  
The 100-pc ring appears to trace the 
system of  stable $x_2$ orbits predicted for the barred Galactic potential. 
\sgra\ is displaced with respect to the geometrical center of symmetry of 
the ring. The ring is twisted and   its morphology suggests a flattening-ratio 
of 2 for the Galactic potential, which is in good agreement with 
the bulge  flattening ratio derived from the 2MASS data. 
\end{abstract}


\section{Introduction}

Observations of the central regions of our Milky Way Galaxy provide the 
nearest  template for the study of the conditions in galactic nuclei in 
normal galaxies. The $3.6\times 10^6$ \msun\ black  hole at the center 
dominates the ÔCentral Molecular ZoneÕ (CMZ)  which  extends from Galactic
longitude l = -1\adeg\ to about  +1.5\adeg\ and hosts the densest
and most massive molecular clouds in the Milky Way \citep{morser96}. The CMZ may be a Galactic analog  of the nuclear star-forming
rings observed in the centers of star forming barred galaxies (\citealt{kk04}, \citealt{kc04}).
The CMZ contains tens of millions of solar masses of cold interstellar matter (ISM) 
\citep{ppetal00,bally10b}  and harbors several of its most active sites of star formation
such as  Sgr A, Sgr B2, and Sgr C (\citealt{yusef08},\citealt{yusef09}).
The gas kinematics  and distribution of near-infrared light indicate
that the gravitational potential in the inner Galaxy is dominated by
a bar with a major axis inclined by about 20 to 45\adeg\ with respect
to our line-of sight (\citealt{binney91}, \citealt{ben03}, \citealt{minchevetal07}).  

Despite the active star formation and bar dynamics,  warm dust has 
been found to be  a relative minor constituent in the CMZ. 
\citet{sod94}  used COBE/DIRBE data to constrain dust properties in 
the CMZ and \citet{rodfer04} presented ISO observations. 
These studies found that 15 to 30\% of the far-IR emission comes
from molecular clouds with dust temperatures of about 19 K, 70 to 75\%
arises from the HI phase with dust temperatures of 17 to 22 K, and
less than 10\% is emitted by warm dust with T $>$ 29 K associated
with  the extended HII and warm (100 to 300 K) molecular phases traced by
H$_3^+$ absorption (\citealt{oka05}, \citealt{goto08}). 
With Herschel we can now extend this type of  analysis into a new 
domain of spatial resolution and sensitivity.

In this Letter, we present the first
high-angular resolution maps of the CMZ far-IR dust emission,
temperature, and column density based on data obtained with the 
Herschel satellite \citep{pilbratt10}.

\begin{figure*}[t]
\resizebox{\hsize}{!}{\includegraphics{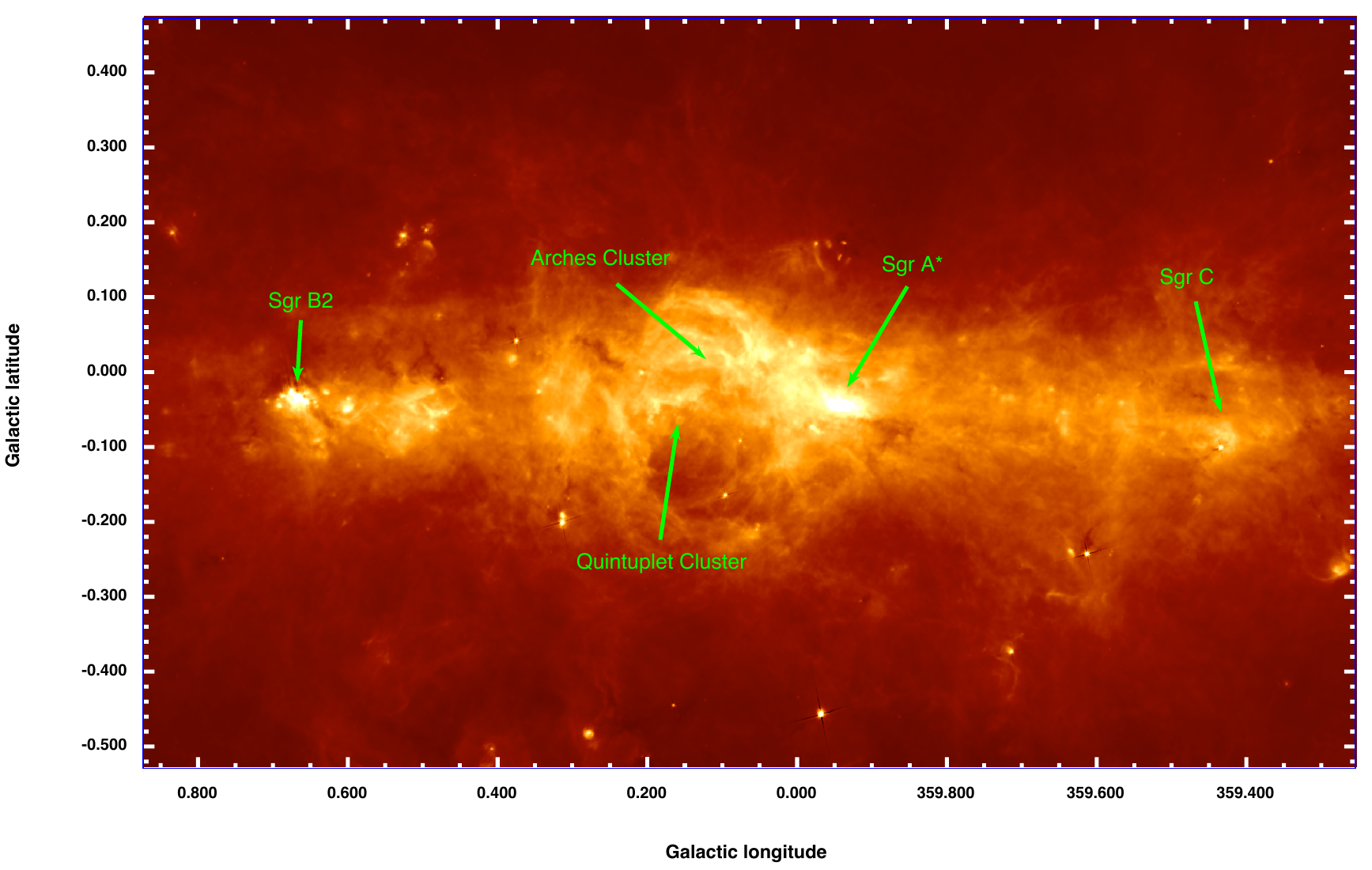}}
\caption{Herschel PACS 70 \um\  image of the Galactic Center region. 
Labels identify known objects that are discussed in the text.}
\label{gc70}
\end{figure*}

\section{Observations and results}

The Galactic Center was observed with  Herschel as part of the Hi-GAL Key-Project 
(the Herschel infrared Galactic Plane Survey, \citealt{moli10a}).  Observing strategy and basic data reduction are described in \citet{moli10b} 
and \citet{trafi11}. The central 2\adeg x2\adeg\ region of the Galaxy was imaged 
in Parallel Mode by acquiring simultaneously PACS \citep{poglitsch10} and 
SPIRE \citep{griffin10} images in 5 photometric bands centered at 70, 160, 250, 
350 and 500 \um. SPIRE was used in "bright-source mode", to minimize non-linearities and 
 saturation in this very bright region; preliminary SPIRE calibration checks indicate an accuracy within 15\%\ compared to standard SPIRE setting (M. Pohlen, priv. comm.). We corrected the zero-level in our SPIRE and PACS images  
using offset values derived from the comparison with the Planck/IRAS data using 
the  same procedure as in \citet{bernard10}. 

\begin{figure*}[t]
\resizebox{\hsize}{!}{\includegraphics{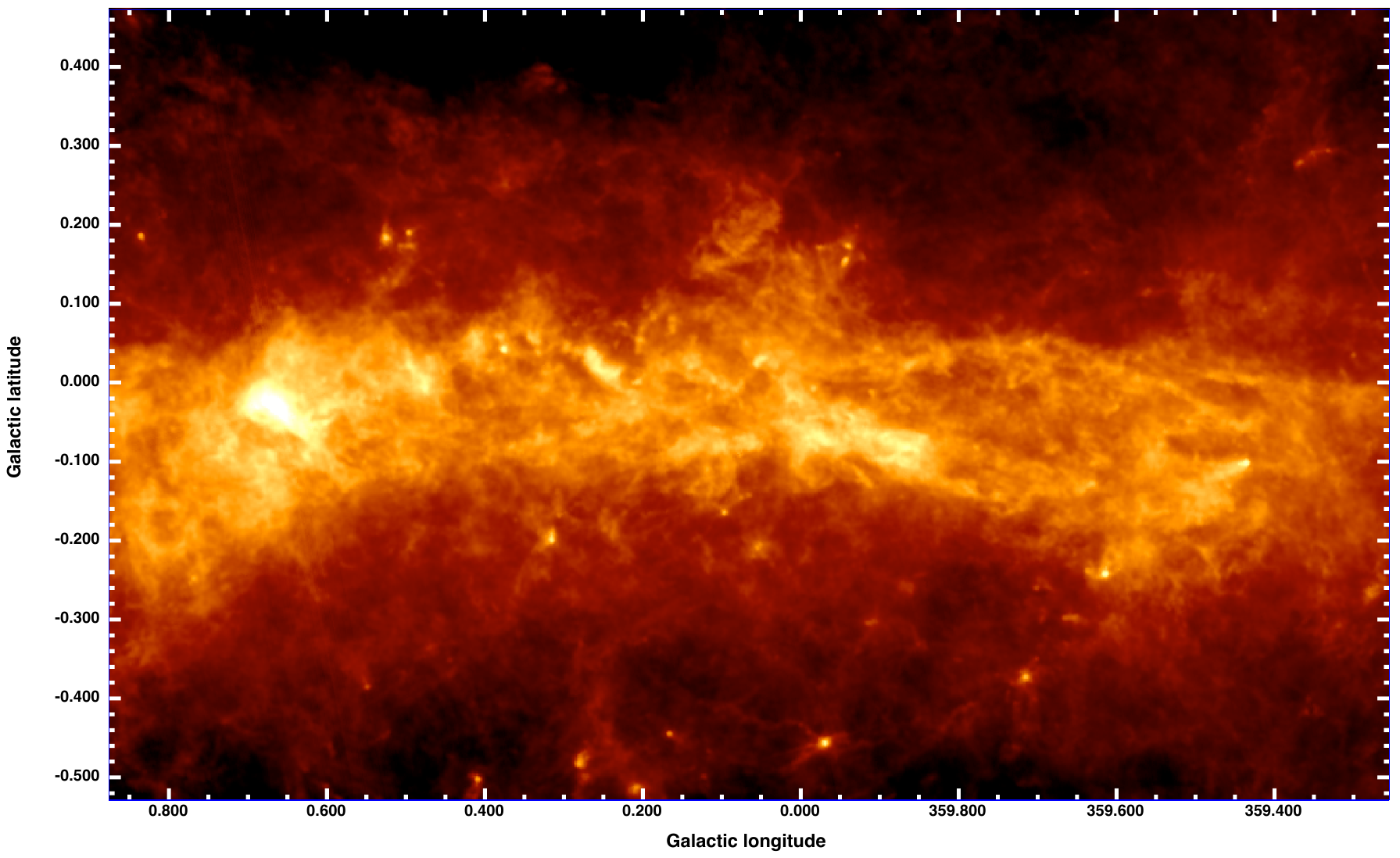}}
\caption{Herschel SPIRE 250 \um\  image of the Galactic Center region.}
\label{gc250}
\end{figure*}

 
Figure \ref{gc70} shows the PACS 70 \um\  image. \sgra\ and \sgrb\ are the brightest spots, and the entire area is filled 
with intense far-IR emission punctuated by several InfraRed Dark Clouds (IRDCs) 
seen in silhouette.  A large bubble dominates at 70 \um\  from 0\adeg $\lsim l \lsim$ +0.20\adeg\  
but  disappears at longer wavelengths (see SPIRE 250 \um\  in fig. \ref{gc250}). 
At 250 \um\  \sgrb\ is  the brightest feature (it is saturated at 250 \um).   Two bent 
chains of emission extend above and below the mid-plane from 
\sgrb\ to \sgra, where they appear to cross, and continue on to \sgrc .   This structure 
can also be identified in the  SCUBA map \citep{ppetal00}, but only the unique multi-band 
far-IR/submm  capabilities of the Herschel PACS and SPIRE cameras make 
possible the determination of its temperature and column density.

Opacity and dust temperature are estimated from a pixel-to-pixel fit to the 
70-350 \um\ data (the resolution is matched to the  350 \um\  images -- $\sim$25\asec) 
using the DustEM model \citep{comp11}  following the method described  
in \citet{bernard10}.    Assuming constant dust properties,   the opacity is
converted into a hydrogen  column density ($N_H\,=\,N_{HI}+N_{H_2}$) using
$\tau_{250}/N_H\,=\,8.8\,10^{-26}$ cm$^{2}$/H.  This value is estimated from
latitudes  b$>$0, and it could be higher by a factor 2 - 4 in the Galactic plane 
\citep{bernard10} and could change our mass estimates accordingly. 
The temperature and column  density maps (Figs. \ref{gc_tmap} and  \ref{gc_nmap}) 
represent the integrated contribution of dust along the line of sight.   
High opacity, cold dust appears in silhouette against the warmer background 
even at 70 \um\  (fig. \ref{gc70}).    Although the derived temperature might be 
underestimated, a temperature map without the 70 \um\ data provides similar results. 

The detailed study of the great bubble visible in the 70 \um\ and the 
temperature maps (figs. \ref{gc70} and \ref{gc_tmap}) will be presented 
in a forthcoming paper. This Letter is  dedicated to the analysis of 
the large-scale filaments visible in the far-IR 
(fig.\ref{gc250}), that trace a continuous chain of cold and dense clumps 
(T$_{dust}\leq$20 K; $N(H) >  2\times 10^{23}$ \cmtwo)  organized along 
an $\infty$-shaped  feature that dominates the image between 
the \sgrc\ complex at  $l$ = 359.4\adeg\  and \sgrb\ at   0.7\adeg. 
The total projected  extent  is $\sim$1.4\adeg, or about 180 pc, for  
a solar Galactocentric distance of 8.4 kpc  \citep{reid09}.  Warmer 
dust with relatively lower column density  
(T$_{dust}\geq$25 K, $ N(H) <  2 \times 10^{23}$ \cmtwo) fills the interior of 
the $\infty$-shaped  feature. 

\section{A 100-parsec elliptical ring of cold and dense molecular clouds: the $x_2$ orbits}

\begin{figure*}[t]
\resizebox{\hsize}{!}{\includegraphics{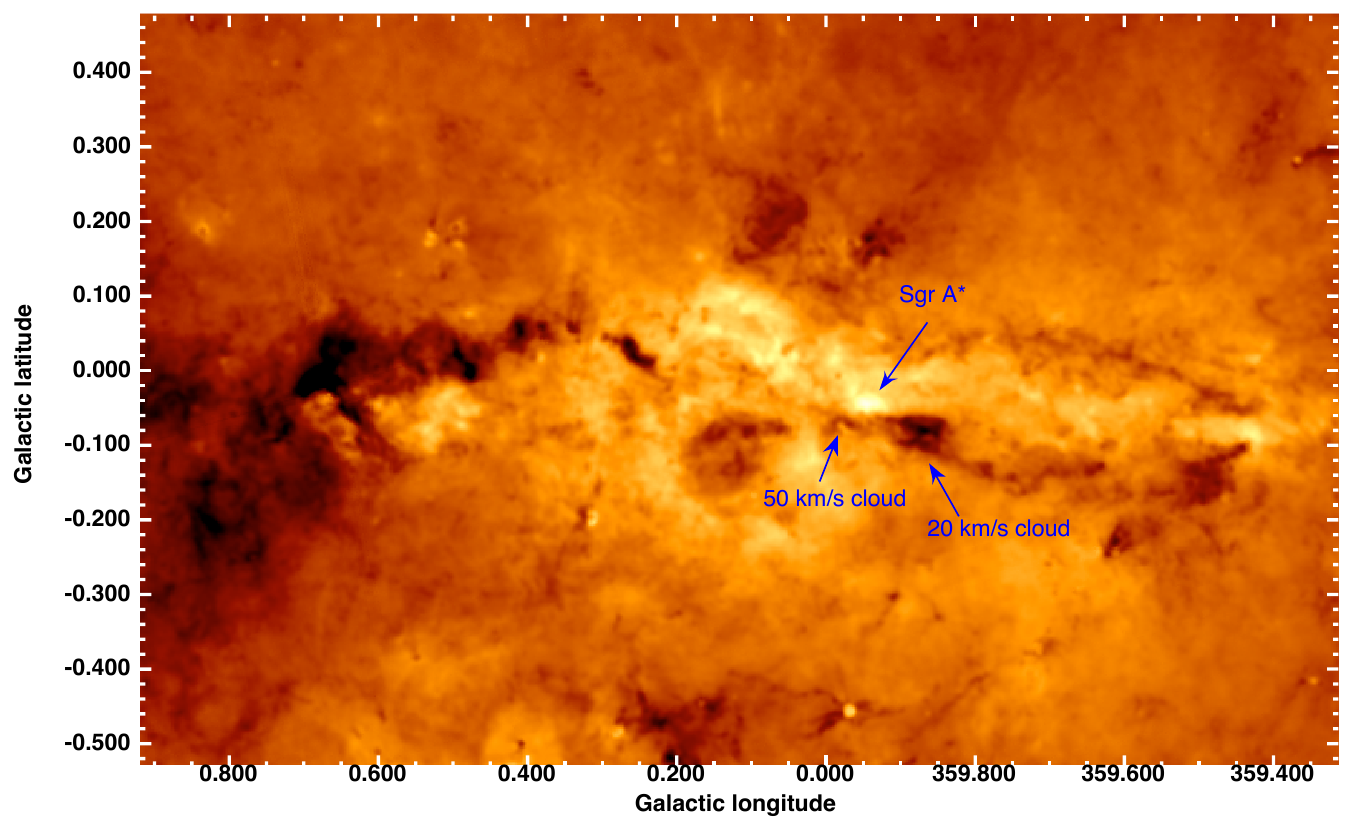}}
\caption{Temperature map of the Galactic Center region. The Log-color stretch extends 
from 15K  to 40K on \sgra.}
\label{gc_tmap}
\end{figure*}

The  $\infty$-shaped feature in the temperature map in fig. \ref{gc_tmap}  coincides with  
high-column density material ($N(H)\geq 2\times 10^{23}$ \cmtwo) organized 
in a continuous chain  of irregular clumps (see fig. \ref{gc_nmap}).    Most of the cold, high column density 
clumps in the \lmbm\ and \lpbp\  arms of this feature are
seen in silhouette against the    70 \um\  background (fig. \ref{gc70}); 
the same is not true for clumps along the \lpbm\ and \lmbp\ arms.  We conclude that 
the \lmbm\ and \lpbp\ arms are in front 
of the bulk of the  warmer dust emission (i.e. between us and 
the Galactic Center), while the \lmbm\ and \lpbp\ arms are located  in 
the background. 

We use the CS survey by 
\citet{tsuboi99} (their fig. 4) to extract the radial velocity of CS condensations positionally coincident with 
the cold and dense peaks on varius positions along the $\infty$-shaped feature in figs. \ref{gc_tmap} and \ref{gc_nmap}. The 
position-velocity information is averaged in latitude bins of 1.5\arcmin, so the 
information can be extracted with some degree of approximation but should 
be adequate to represent the mean velocity of the various clouds and clumps. 
The full span of velocities found toward each position is reported in fig. \ref{gc_nmap}. 

The \lmbp\ arm shows steadily increasing (in absolute values) negative velocities 
 from the center of symmetry toward  \sgrc\, indicating 
approaching material with increasing relative speed. Following the \lmbm\ 
arm from \sgrc\ back to the centre we find negative velocities decreasing and 
changing to positive (receding) values  at $l\sim$359.8\adeg. Continuing 
along the \lpbp\ arm the radial velocities are still positive and slowly increasing, 
indicating material receding at higher and higher speeds up to \sgrb. Finally along the \lpbm\ arm we see velocities decrease 
again going back toward the center of symmetry of the $\infty$-shaped feature. 
The distribution of radial velocities along the $\infty$-shaped 
feature seems to indicate an ordered, rotating pattern around its projected 
center of symmetry. 

\subsection{A simple model for the $\infty$-shaped feature: a 100-pc elliptical twisted ring}

A simplified toy model of the $\infty$-shaped feature  is built assuming that 
the material is distributed along an elliptical orbit (projected onto 
the Galactic Plane) with semi-axes [$a,b$] and major-axis position angle $\theta_p$. 
For simplicity,  a constant  orbital speed \textrm{v}$_{orb}$  is assumed. 
An additional sinusoidal vertical 
oscillation component with a vertical frequency $\nu_z$ and a phase $\theta _z$ is added
to the equations of the ellipse describing the orbit.    Adopting a coordinate 
system with the $x$ axis oriented toward negative longitudes, $y$ axis pointing 
away along the line of sight from the Sun, and $z$ axes toward positive latitudes 
(see Fig. \ref{ring}), the position and radial velocity of each point along the orbit 
is described as a function of the polar angle $\theta_t$ (counter-clockwise
from the  $x$ axes) as 

\begin{displaymath}
\left\{ \begin{array}{l}
x=a\,\cos \theta_t \, \cos \theta_p - b\, \sin \theta_t \, \sin \theta_p \\
y=a\, \cos \theta_t \, \sin \theta_p + b\, \sin \theta_t \, \cos \theta_p \\
z=z_0\, \sin \nu_z (\theta_p - \theta_z)\\
\textrm{v}_r=-\textrm{v}_{orb} \, \cos(\theta_p +\theta_t)
\end{array} \right.
\label{ringmodel}
\end{displaymath}

\begin{figure*}[t]
\resizebox{\hsize}{!}{\includegraphics{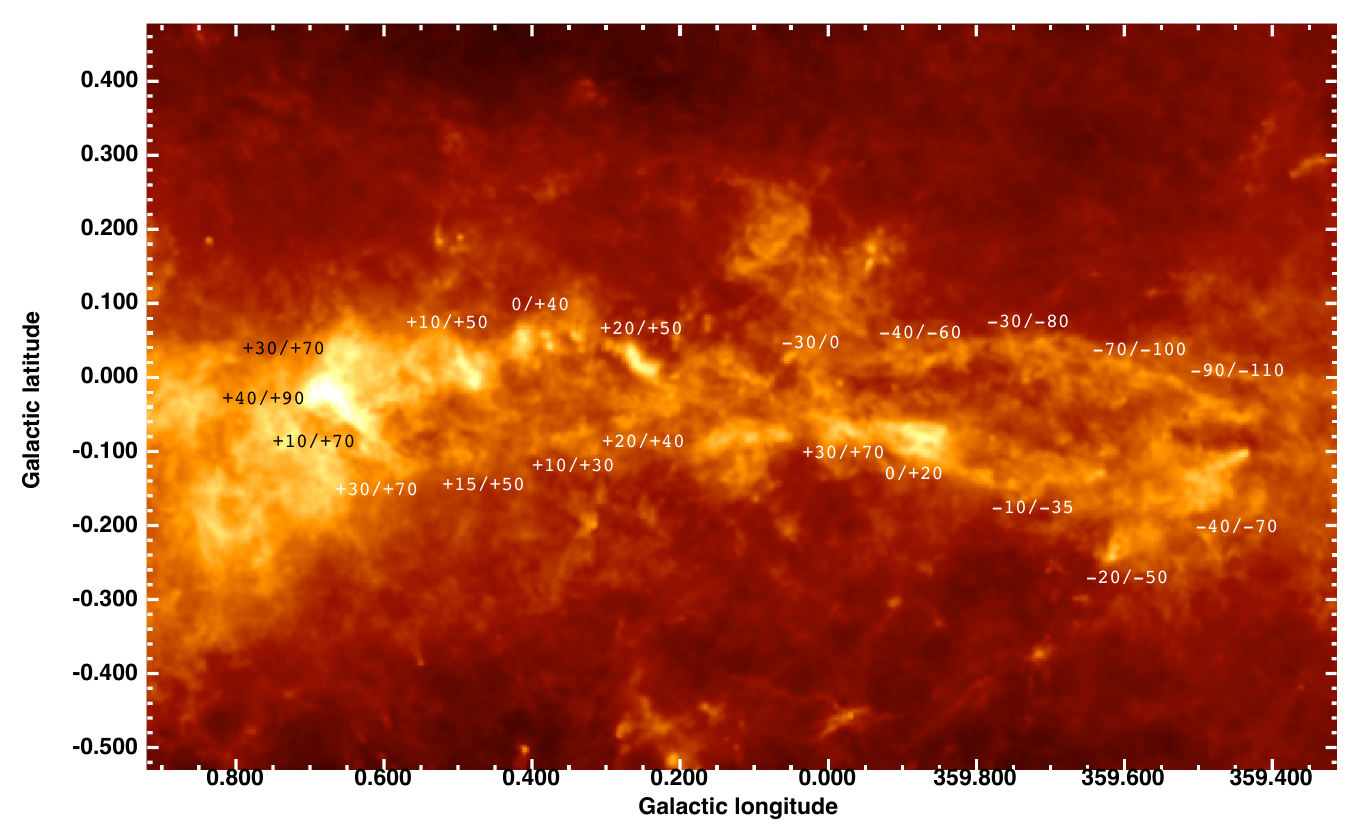}}
\caption{Atomic hydrogen column density map of the Galactic Center region. 
Color scale shown is logarithmic and extends from  
$4\times 10^{22}$ \cmtwo\ in the dark regions, to $4\times 10^{25}$ \cmtwo\ 
in the brightest region corresponding to \sgrb. Velocity information is extracted 
from CS spectroscopic cubes \citep{tsuboi99} for the gas counterparts positionally 
associated with the dense dust clumps; the range values indicate the range where 
emission is observed.
}
\label{gc_nmap}
\end{figure*}

\begin{figure*}[t]
\resizebox{\hsize}{!}{\includegraphics{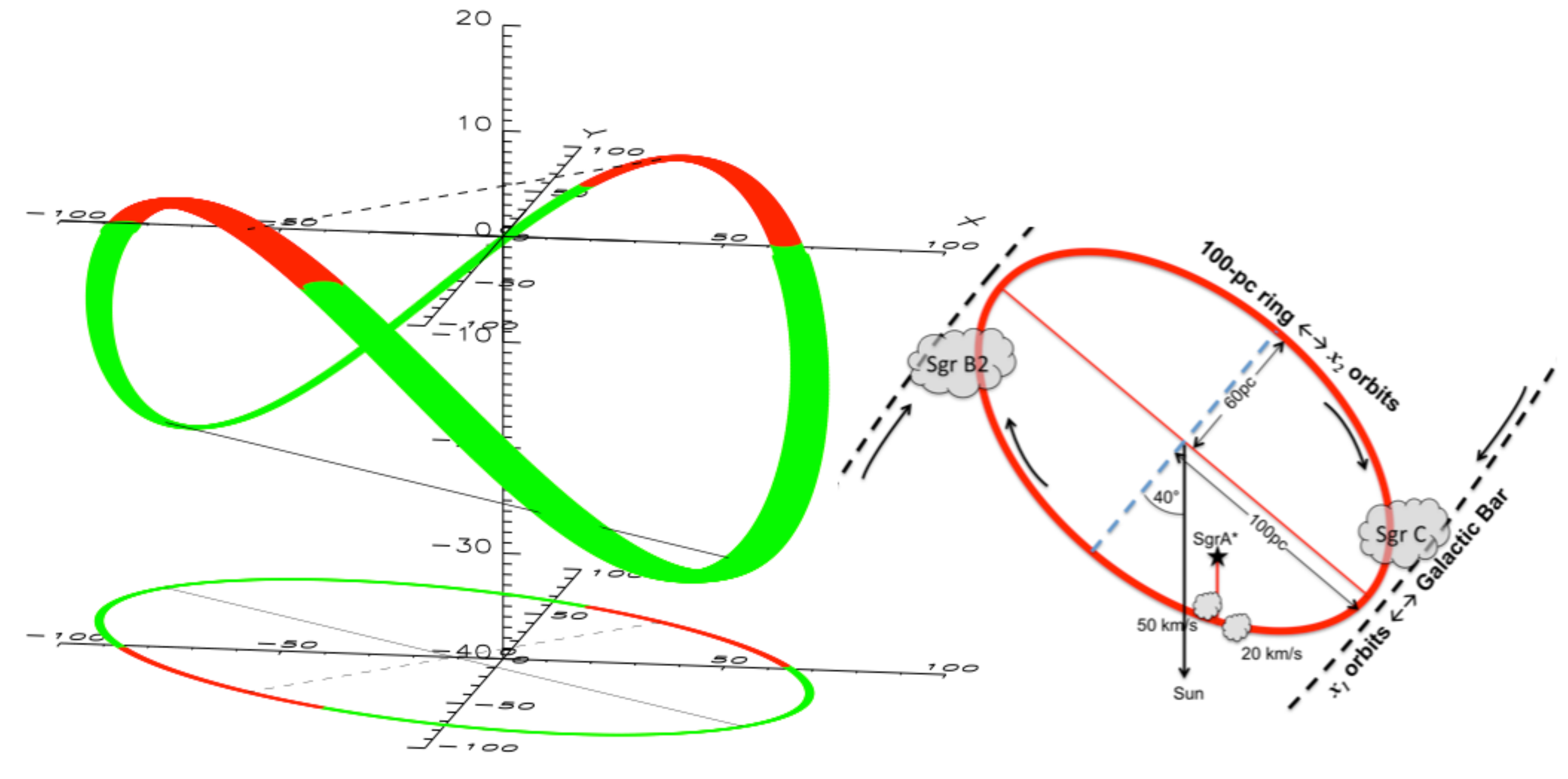}}
\caption{Sketch of the proposed 3D structure and placement of the 100-pc ring. \textit{Left panel:} the ring is represented by the thick color line (red and green mark positions above and below \textit{b}=0\adeg. The line of sight to the Sun is along the Y axes. The thin full and dashed lines represent the major and minor axes of the ellipse. \textit{Right panel:} Top view of the 100-pc ring with the proposed location of major clouds. The thick black dashed lines represent the innermost $x_1$ orbits. The position of SgrA* along the line of sight is the one corresponding to the distance it should have from the front portion of the 100-pc ring to justify the velocity difference between the 20 and 50 \kms\ clouds if due to the gravitational pull of the mass concentration around \sgra\ (see text).}
\label{ring}
\end{figure*}

We let $a, b, \theta_p$ and \textrm{v}$_{orb}$ vary until they 
match the observed projected morphology of the ellipse and the distribution of 
radial velocities at 20 positions along the $\infty$-shaped feature.  We keep $z_0$ fixed at 15 pc, 
which is half of the measured $b$-extent of the $\infty$-shaped feature;  we also fix $\nu_z$=2 in units of the orbital frequency, as 
indicated by the projected  twist shape (see \S\ref{twist_par}).    The fitting procedure minimizes a 
pseudo-$\chi^2$ value $\xi$

\begin{equation}
\xi = \sum _{i=1} ^{20} {{(\textrm{v}_r-\textrm{v}_{obs})^2}\over{\Delta \textrm{v}_{obs}}}
\end{equation}

where $\Delta \textrm{v}_{obs}$ is half of the  velocity range toward 
each of the 20 test positions (see fig. \ref{gc_nmap}). 

The best fit is obtained for an ellipse with $a=100$pc, $b=$60pc, $\theta_p=-$40\adeg, and   \textrm{v}$_{orb}$=80 \kms ; $\theta _z \sim$0 implying that the ring crosses the midplane along the X axes. These numbers should be taken with caution, as the velocities are derived by eye from a plot. However, the discrepancy between the model radial velocity and 
the observed ones is within $\pm$25 \kms. The only  exception is the  
50 \kms\ cloud (see fig. \ref{gc_tmap}) located  close in projection to \sgra\ (see \S\ref{sgradist}).

The CMZ has been the subject of many studies in millimeter molecular tracers, and although the interpretation is difficult for the extreme confusion along the line of sight, several  features were identified. \citet{sofue95} analysed the 
CO data in \citet{bally87} and identified two large-scale structures (the "Arms"). \citet{tsuboi99} 
recognized a  large-scale feature in their CS position-velocity diagram, that they call 
"Galactic Center Bow"  which matches the \lmbm -\lpbp\ arms of the $\infty$-shaped feature. The cold, high column-density dust features identified in the Herschel images, used as a proxy for the densest molecular gas, now indicate that the
previously identified "Arms" and the "Galactic Center Bow" are parts of a
single, twisted and elliptical ring rotating around \sgra.
Figure \ref{ring} presents a sketch of the proposed arrangement of what we will call the100-pc ring. 

A rough estimate of the ring mass is obtained by integrating the N(H) map 
within a contour surrounding the 100-pc ring  which encloses material with 
N(H)$\gsim 2\times 10^{23}$ \cmtwo\  (for a corresponding solid angle 
of $\sim 4\times 10^{-5} sr$).   The contribution of  background or foreground 
gas is removed by subtracting a  column density of N(H) = $2\times 10^{23}$ \cmtwo, 
the mean N(H) value in the interior of $\infty$-shaped feature.   The resulting ring 
mass is  $\sim 3\times 10^7$ \msun\ (it is $\sim 4\times 10^7$ \msun\ 
without background subtraction).    Within the uncertainties, the derived 
ring mass agrees with  5.3$\times 10^7$ \msun\  estimated 
from SCUBA maps \citep{ppetal00} and  3$\times 10^7$ \msun\ 
from CO data \citep{dahmen98}. 

The 100-pc ring traces gas and dust moving about  the nucleus on an 
elliptical orbit whose major axis (the straight red line in the right panel of fig. \ref{ring}) is 
perpendicular to the major axis of the great Galactic bar, oriented 20 to 45\adeg\ 
with respect to our line-of-sight (see fig. \ref{ring}). 
The 100-pc ring is likely located on the so-called `$x_2$'  orbit system, which precesses at the same 
rate as the solid-body  angular velocity of the Galactic Bar, resulting in stable, 
non-intersecting  trajectories  \citep{binney91}. These $x_2$ orbits are enclosed 
within another larger  system of stable  orbits, called $x_1$,  elongated  along the Galactic Bar. 
The prominent  \sgrb\ and \sgrc\  star-forming complexes are located at 
the two projected longitude extrema of the 100-pc ring where gas on $x_1$ and
$x_2$ orbits may collide, producing strong shocks which may trigger the formation of massive molecular cloud formation as well as enhanced star formation rates by the shock focusing mechanism 
suggested by \citet{kl91} for the spiral-bar interface of M83. 

It is tempting to speculate that the 100-pc ring  constitutes the remnant of 
a more persistent dusty torus that may have played a role in  past AGN phases 
of our Galaxy. The recent Fermi-LAT detection of a large gamma-ray bubble 
emanating from the Galactic Center (\citealt{suetal10}, \citealt{croker11}, \citealt{zubo11}) 
provides possible evidence for such past activity. Mid-IR interferometry  
 of AGNs  reveals compact dusty  structures with a 
radius of a few parsecs,  (e.g. \citealt{jaffeetal04}, \citealt{radom08}), but there 
is additional evidence  from Spitzer imaging and spectroscopy suggesting 
that larger and more  persistent structures play a role in shaping 
observed AGN properties  (e.g. \citealt{shietal06}, \citealt{quill06}).   Model 
predictions (e.g. \citealt{krolik07})  for torus column density in 
excess of 10$^{24}$ \cmtwo,  and orbital speed $\sim$100 \kms, seem to be 
consistent with the parameters we estimate for the 100-pc ring.

\subsection{The relation between the 100-pc ring and \sgra}
\label{sgradist}

\sgra  is noticeably displaced $\sim$10\amin\ (24 pc in projection) towards 
the negative longitude side of the geometrical center of symmetry of the 100-pc ring. 
The ring displacement  with respect to the supposed location of the Galactic Center  
is consistent  with the overall displacement of dense gas within the CMZ. 
Two-thirds to  three-quarters of the cold  interstellar gas and dust  within the inner 
few  hundred parsecs of the nucleus is at positive longitudes  with respect 
to \sgra (\citealt{bally88}, \citealt{morser96}).  A similar asymmetry is present
in the observed radial velocity  on the 100-pc ring close to the location 
of \sgrb, as opposed to \sgrc\ (see fig. \ref{gc_nmap}). This  may 
indicate an additional pattern motion of $\sim 10-20$ \kms\  of the 100-pc 
ring as a whole along its major  axes and toward us, implying a `sloshing' motion  
back and forth that 
could be explained as the response of the gas to an m=1  mode in the  
Galactic bar potential \citep{morser96}.  



The  projected proximity of the 20 and 50 \kms\  clouds to \sgra, 
has led to the suggestion (e.g. \citealt{herrho05})  that they are 
physically interacting with \sgra . 
If they are part of the 100-pc ring, then \sgra\ must be closer to 
the front of the ring than to its back;  \sgra\  may then not be in  the ring center,  
nor lie along  the ellipses axes.   Proximity to \sgra\  could be the source  of the 
roughly 40 \kms\ deviation of the radial velocity of  the 50 \kms\ cloud from the 
value predicted by the  model of the rotating elliptical ring. 

Given the extent of the 100-pc ring, its orbital motion is driven by the enclosed gas and bulge stars. If, however, ring material passes in proximity of \sgra\ the gravitational pull due to the strong mass concentration can be substantial. Indeed, the $3.6\times 10^6$ \msun\ black hole in \sgra\ dominates the gravitational
potential only to a radius of a few parsecs.   Beyond that distance the potential
is dominated by the stellar mass in the  bulge with a radial density profile
of the form $\rho (r) \approx \rho_0 (r/r_b)^{-\alpha}$ where $\alpha \approx -1.9$, $r_b\sim 0.34$pc and $\rho_0 \sim 2.1\times 10^6$ \msun/pc$^3$ \citep{trippe08}. We then treat the gravitational pull from the \sgra\ mass concentration as an additional component over the general rotational motion. Using Newton's 2$^{nd}$ law:

\begin{equation}
{{GM_C(r)}\over{r^2}}={{\Delta \textrm{v}}\over{t_0}}
\label{fma}
\end{equation}

where $r$ is the distance between \sgra\ and the two molecular clouds (that is assumed similar for the two clouds if they belong to the 100-pc ring), $M_C(r)$ is the mass enclosed in a sphere of radius $r$ centered on \sgra\, and t$_0$ is the timescale for the cloud passing in front of \sgra. The latter can be written as $t_0=\Delta x / \textrm{v}_{orb}$, where $\Delta x$ is the projected distance between the 50 and 20 \kms\ clouds and $v_{orb}$ is their orbital speed. Eq. \ref{fma} can then be rewritten

\begin{equation}
r=\sqrt{{GM_C(r) \Delta x}\over{\Delta \textrm{v}\, \textrm{v}_{orb}}}
\label{rdiff}
\end{equation}

Using this formulation to estimate M$_C(r)$, the measured projected distance between the nearest edges of the two clouds of $\Delta x \sim 7$pc, and assuming an orbital speed of 80\kms\ from our toy model fit, Eq. \ref{rdiff} provides $r\sim $20 pc. In other words, to explain the difference in radial velocities between the 20 and 50 \kms clouds would require that \sgra\ is closer than 20pc to the foregreound arm of the 100-pc ring.

\subsection{The ring's twist: constraints on the flattening of the Galactic potential}
\label{twist_par}

Perhaps the most striking aspect of the 100-pc ring is its twist.  Apparently 
the gas in the ring oscillates twice about the mid-plane each time it orbits 
the nucleus once. Assuming the elliptical axes and the orbital speed derived 
from the toy model,  the orbital period is 
P$_o \sim 3 \times 10^6$ years; the period of the vertical oscillations 
is therefore P$_z \sim 1.5 \times 10^6$ years. The coherence of the ring's 
morphology suggests  stability over a few revolutions. 
Thus, the relationship between P$_o$ and P$_z$ can be used to 
infer constraints on the flattening of the gravitational potential in the CMZ. 
Assuming as a gross approximation that the matter (gas and stars) 
is distributed in a uniform-density $\rho _0$ slab of thickness $h$, P$_o$ 
at a distance R can then be expressed as 

\begin{equation}
P_o=\sqrt{{4\pi R}\over{G \rho_0 h}}
\label{per_orb}
\end{equation}

A perturbation on a particle rotating in the flattened slab potential will result, 
in a 1$^{st}$-order approximation \citep{bt}, in a harmonic oscillation both in 
radial and vertical directions. The vertical oscillation frequency 
can be expressed as 

\begin{equation}
\nu _z ^2 = \left({\partial ^2 \Phi}\over{\partial z^2} \right)_{R,0}
\label{nu_z}
\end{equation}

The gravitational field along the $z$ direction of the  slab 
(approximated as infinite) is 

\begin{equation}
F _z = 2\pi \rho_0 G h 
\label{f_z}
\end{equation}

so that the period of the vertical oscillations is

\begin{equation}
P _z  = \sqrt{{2 \pi}\over{\rho_ 0 G} }
\label{per_ver}
\end{equation}

As noted above, the $\infty$ shape that the ring displays in projection implies that

\begin{equation}
P _o = 2 P_z
\end{equation}

or, using Eqs. \ref{per_orb} and \ref{per_ver}, 

\begin{equation}
h={{R}\over{2}}
\label{hr}
\end{equation}

The shape of the 100-pc ring would therefore imply a flattened potential in the 
CMZ that is consistent with a $x/z=0.5$ axial ratio bulge mass distribution, 
remarkably similar to the value of 0.5$-$0.6 recently derived \citep{rc08} 
fitting the 2MASS data to a potential model that includes the bulge and the nuclear bar.

\section{Conclusions}
The far-infrared multi-band images provided by the Hi-GAL survey using 
the Herschel PACS and SPIRE cameras provide evidence for a 
100x60 parsecs elliptical ring of cold and dense clouds orbiting around the 
Galactic Center on $x_2$ orbits elongated along the minor axis of 
the Galactic center Bar. The ring is twisted with a vertical frequency that is twice the orbit frequency,  
indicating  a flattening-ratio of 2 for the Galactic Center gravitational
potential, consistent with the value  derived from bulge star 2MASS 
number counts.
\acknowledgments
Data processing has been possible thanks to 
ASI support via contract I/038/080/0. We thank the referee, M. Morris, whose comments led to a more focused paper. We thank L. Spinoglio, N. Sacchi and A. Marconi for useful 
discussions on AGNs, and M. Pohlen and the SPIRE ICC for the  preliminary
analysis of the SPIRE "bright source mode" calibration.


\end{document}